\documentclass[ALICE,manyauthors]{cernphprep}

\usepackage[comma,square,numbers,sort&compress]{natbib}
\usepackage{hyperref}

\newcommand{\He}{$\rm{^{3}He}$~}

\newcommand{\Hens}{$\rm{^{3}He}$}

\newcommand{\Hebar}{$\rm{^{3}\overline{He}}$~}

\newcommand{\Hebarns}{$\rm{^{3}\overline{He}}$}

\newcommand{\deutns}{$\rm{d}$}
\newcommand{\dbar}{$\rm{\overline{d}}$~}
\newcommand{\dbarns}{$\rm{\overline{d}}$}

\begin{document}%

\begin{titlepage}
\PHyear{2015}
\PHnumber{035}      
\PHdate{19 February}  
%

\title{Precision measurement of the mass difference\\between light nuclei and anti-nuclei}
\ShortTitle{Precision measurement of the mass difference between light nuclei and anti-nuclei}   

\Collaboration{ALICE Collaboration\thanks{See Appendix~\ref{app:collab} for the list of collaboration members}}
\ShortAuthor{ALICE Collaboration} 

\begin{abstract}
The measurement of the mass differences for systems bound by the strong force
has reached a very high precision with protons and anti-protons.
The extension of such measurement from \mbox{(anti-)baryons} to \mbox{(anti-)nuclei} allows one
to probe any difference in the interactions between nucleons and anti-nucleons encoded
in the \mbox{(anti-)nuclei} masses. This force is a remnant of the underlying strong interaction
among quarks and gluons and can be described by effective theories,
but cannot yet be directly derived from quantum chromodynamics.
Here we report a measurement of the difference between the ratios
of the mass and charge of deuterons (d) and anti-deuterons (\dbarns), and \He and \Hebar
nuclei carried out with the ALICE (A Large Ion Collider Experiment)
detector in Pb-Pb collisions at a centre-of-mass energy per nucleon pair of 2.76 TeV.
Our direct measurement of the mass-over-charge differences confirm CPT invariance
to an unprecedented precision in the sector of light nuclei.
This fundamental symmetry of nature, which exchanges particles with anti-particles,
implies that all physics laws are the same under the simultaneous reversal of charge(s) (charge conjugation C),
reflection of spatial coordinates (parity transformation P) and time inversion (T).

\end{abstract}

\end{titlepage}
\setcounter{page}{2}

The measurement of the mass differences for systems bound by the strong force
has reached a very high precision with protons and \mbox{anti-protons}\cite{pbarMass,TRAP}.
The extension of such measurement from \mbox{(anti-)baryons} to \mbox{(anti-)nuclei} allows one
to probe any difference in the interactions between nucleons and anti-nucleons encoded
in the (anti-)nuclei masses. This force is a remnant of the underlying strong interaction
among quarks and gluons and can be described by effective theories\cite{NuclForces},
but cannot yet be directly derived from quantum chromodynamics.
Here we report a measurement of the difference between the ratios
of the mass and charge of deuterons (d) and anti-deuterons (\dbarns), and \He and \Hebar
nuclei carried out with the ALICE (A Large Ion Collider Experiment)\cite{ALICE}
detector in Pb-Pb collisions at a centre-of-mass energy per nucleon pair of 2.76 TeV.
Our direct measurement of the mass-over-charge differences confirm CPT invariance
to an unprecedented precision in the sector of light nuclei\cite{CPTtheorem_1,CPTtheorem_2}.
This fundamental symmetry of nature, which exchanges particles with anti-particles,
implies that all physics laws are the same under the simultaneous reversal of charge(s) (charge conjugation C),
reflection of spatial coordinates (parity transformation P) and time inversion (T).

Heavy ions are collided at very high energies at the CERN Large Hadron
Collider (LHC) to study matter at extremely high temperatures and densities.
Under these conditions heavy-ion collisions are a copious source of matter
and anti-matter particles and thus are suitable for an experimental
investigation of their properties such as mass and electric charge. In
relativistic heavy-ion collisions, nuclei and corresponding
anti-nuclei are produced with nearly equal rates\cite{STARprod}.
Their yields have been measured at the Relativistic Heavy Ion Collider (RHIC)
by the STAR\cite{STAR} and PHENIX\cite{PHENIX} experiments and at the
LHC by the ALICE\cite{ALICE} experiment.
To date, the heaviest anti-nucleus which has been observed\cite{STARprod}
is $\rm{^4\overline{He}}$ (anti-$\rm{\alpha}$); meanwhile, for lighter nuclei and anti-nuclei, which are more copiously produced,
a detailed comparison of their properties is possible.
This comparison represents an interesting test
of CPT symmetry in an analogous way as done for elementary fermions\cite{posmass,gFactor} and bosons\cite{dm_bosons}, and for QED\cite{ALPHA2012,Hbarcharge} and
QCD systems\cite{pbarMass,TRAP,nbarMass,nbarMassErrata,ATRAP} (a particular example for the latter being the measurements carried out on neutral kaon
decays\cite{k0CPT}), with different levels of precision which span over several
orders of magnitude.
All these measurements can be used to constrain, for different interactions,
the parameters of effective field theories that add explicit CPT violating terms to the Standard Model Lagrangian,
such as the Standard Model Extension\cite{sme} (SME).

The measurements reported in this paper are based on the high-precision tracking and identification
capabilities of the ALICE experiment\cite{ALICEperf}.
The main detectors employed in this analysis are the ITS\cite{ITSperf}
(Inner Tracking System) for the determination of the interaction vertex,
the TPC\cite{TPCperf} (Time Projection Chamber) for tracking and specific energy loss
(${\rm d}E/{\rm d}x$) measurements, and the TOF\cite{TOFperf} (Time Of Flight) detector to measure
the time $t_{{\rm TOF}}$ needed by each track to traverse the detector.
The combined ITS and TPC information is used to determine the track length ($L$) and the
rigidity ($p/z$, where $p$ is the momentum and $z$ the electric charge in units of the elementary charge $e$) of the charged particles 
in the solenoidal 0.5~T magnetic field
of the ALICE central barrel (pseudo-rapidity $|\eta|< 0.8$).
Based on these measurements, we can extract the squared mass-over-charge ratio
$\mu^2_{\rm{TOF}} \equiv \left ( m/z \right )^{2}_{{\rm TOF}} = (p/z)^2 \, [(t_{\rm TOF} / L)^2 - 1/c^2]$.
The choice of this variable is motivated by the fact that $\mu^2$ is directly proportional to the square of the time of flight, allowing to better preserve its Gaussian behaviour.

\begin{figure}
\begin{center}
\includegraphics[width=0.95\textwidth]{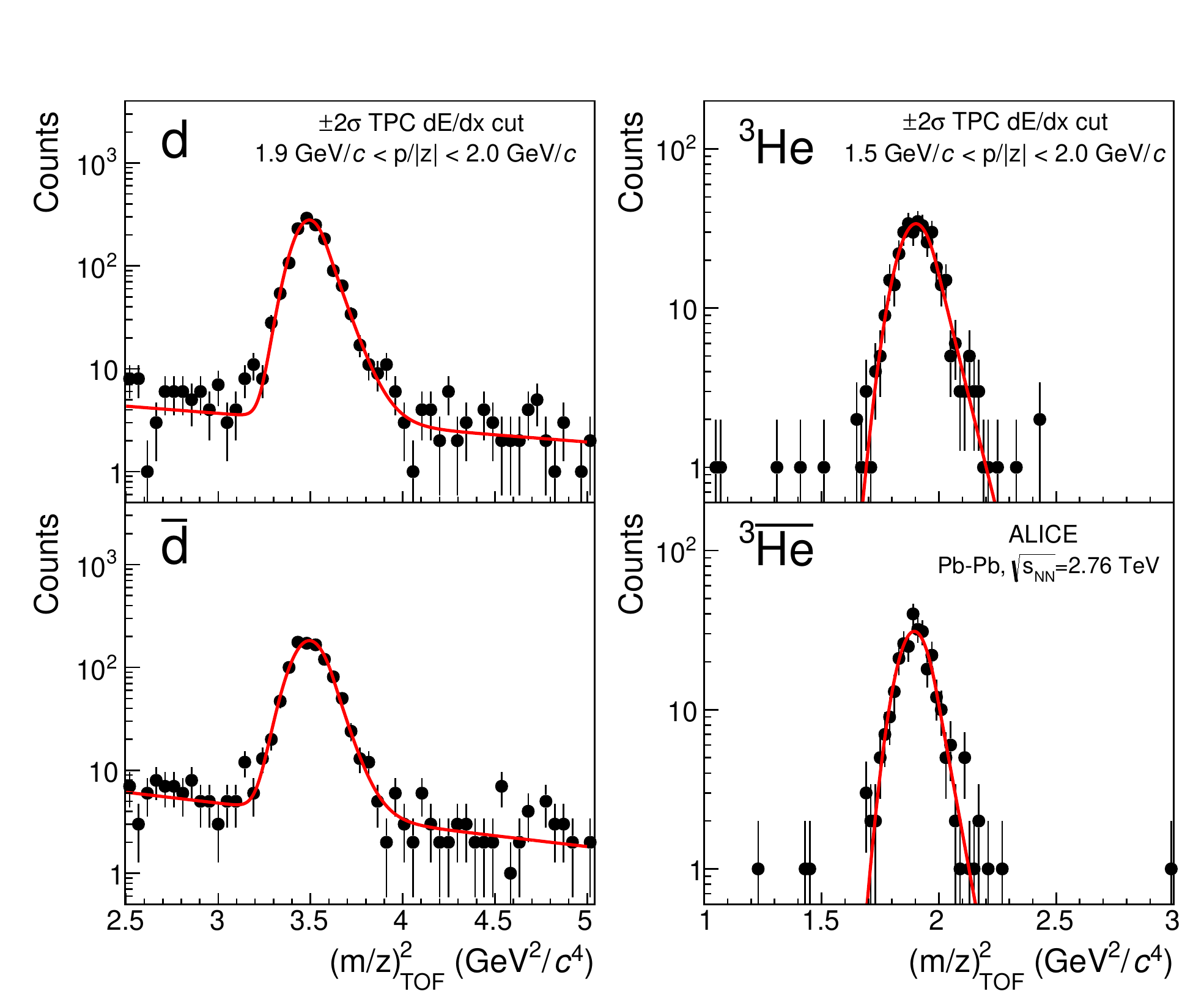}
\caption{Examples of squared mass-over-charge ratio distributions
for deuterons (left) and \He (right) in selected
rigidity intervals. Particle and anti-particle spectra
are in the top and bottom plots, respectively. The fit function (red curve)
also includes, for the (anti-)deuteron case, an exponential term to describe the background.
In the rigidity intervals shown here the background is about 4\% for
(anti-)deuterons, while it is 0.7\% for \He and \Hebar.}
\label{fig:massSpectra}
\end{center}
\end{figure}

The high precision of the TOF detector, which determines the arrival time
of the particle with a resolution of 80 ps\cite{ALICEperf}, allows us to
measure a clear signal for (anti-)protons, (anti-)deuterons and (anti-)\He
nuclei over a wide rigidity range \mbox{($1 < p/|z| < 4~ \rm{GeV}/{\it c}$)}.
The main source of background, which is potentially of the same order of the signal,
arises from tracks erroneously associated to a TOF hit.
To reduce this contamination,
a $2\sigma$ cut (where $\sigma$ is the standard deviation) around the
expected TPC ${\rm d}E/{\rm d}x$ signal is applied.
Such a requirement strongly suppresses (to below 4\%)
this background
for rigidities below $p/|z| <2.0~\rm{GeV/{\it c}}$ for (anti-)deuterons
and for all rigidities for (anti-)\He (to below 1\%).
For each of the species under study, the mass is extracted by
fitting the mass-squared
distributions in narrow $p/|z|$ and $\eta$ intervals, using a Gaussian with a small exponential tail that
reflects the time signal distribution of the TOF detector.
Examples of the mass-squared distributions for (anti-)deuterons and (anti-)\He
candidates are reported in Fig.~\ref{fig:massSpectra} in selected rigidity intervals.

Using mass differences, rather than absolute masses,
allows us to reduce the systematic uncertainties related to tracking, spatial
alignment (affecting the measurement of the track momentum and length) and time calibration.
Despite that, residual effects are still present, due to imperfections in the detector alignment and the description of the 
magnetic field,
which can
lead to position-dependent systematic uncertainties.
In terms of relative uncertainties, the ones affecting the measurement of the momentum are the largest and independent of the mass,
and are the same
for all positive (negative) particles in a given momentum interval.
It is therefore possible to correct the (anti-)deuteron and the (anti-)\He
masses by scaling them with the ratio between the (anti-)proton masses recommended by PDG\cite{PDG} ($\mu_{{\rm p(\bar{p})}}^{{\rm PDG}}$)
and the ones measured in the analysis presented here ($\mu_{{\rm p(\bar{p})}}^{{\rm TOF}}$), i.e.
$\mu_{{\rm A(\bar{A})}}=\mu_{{\rm A(\bar{A})}}^{{\rm TOF}}\times (\mu_{{\rm p(\bar{p})}}^{{\rm PDG}} / \mu_{{\rm p(\bar{p})}}^{{\rm TOF}})$.
These correction factors, which depend on the rigidity, deviate from unity by at most 1\%.
Conversely, systematic effects connected to the track-length measurement are mass dependent and cannot be completely
accounted  for using the above correction.
However, they are expected to be symmetric for positive and
negative particles when inverting the magnetic field.
Any residual asymmetry is therefore indicative of remaining systematic uncertainties related to the
detector conditions.
In order to estimate them and keep these effects under control, both nuclei and anti-nuclei measurements are performed for two opposite magnetic field configurations and then averaged.
Their half difference is taken as the estimate of this systematic uncertainty. 
Other sources of systematic uncertainties are evaluated by varying energy loss
corrections applied to the reconstructed momentum, the range and the shape of the background function assumed in the fit of the mass-squared distributions
and the track selection criteria.
In particular, TPC ${\rm d}E/{\rm d}x$ cuts
are varied between one and four standard deviations
to probe the sensitivity of the fit results on the residual background,
and a tracking quality cut on the distance of closest approach of the track to the vertex is varied to evaluate
the influence of secondary particles on the measurement.
The sources of systematic uncertainties are found to be fully correlated among all the rigidity intervals,
except for those due to the fit procedure and the TPC selection criteria where the uncertainties are uncorrelated.
For deuterons and anti-deuterons, the largest relative systematic uncertainties on $\Delta\mu/\mu$ come from the detector alignment ($\sim 0.7\times 10^{-4}$),
the TPC selection criteria ($\sim 0.7\times 10^{-4}$) and the secondaries ($\sim 1.0\times 10^{-4}$). For \He and \Hebarns,
they come from the energy loss corrections ($\sim 0.7\times 10^{-3}$), the fit procedure ($\sim 0.5\times 10^{-3}$) and the TPC selection criteria ($\sim 0.4\times 10^{-3}$).

\begin{figure}
\begin{center}
\includegraphics[width=0.65\textwidth]{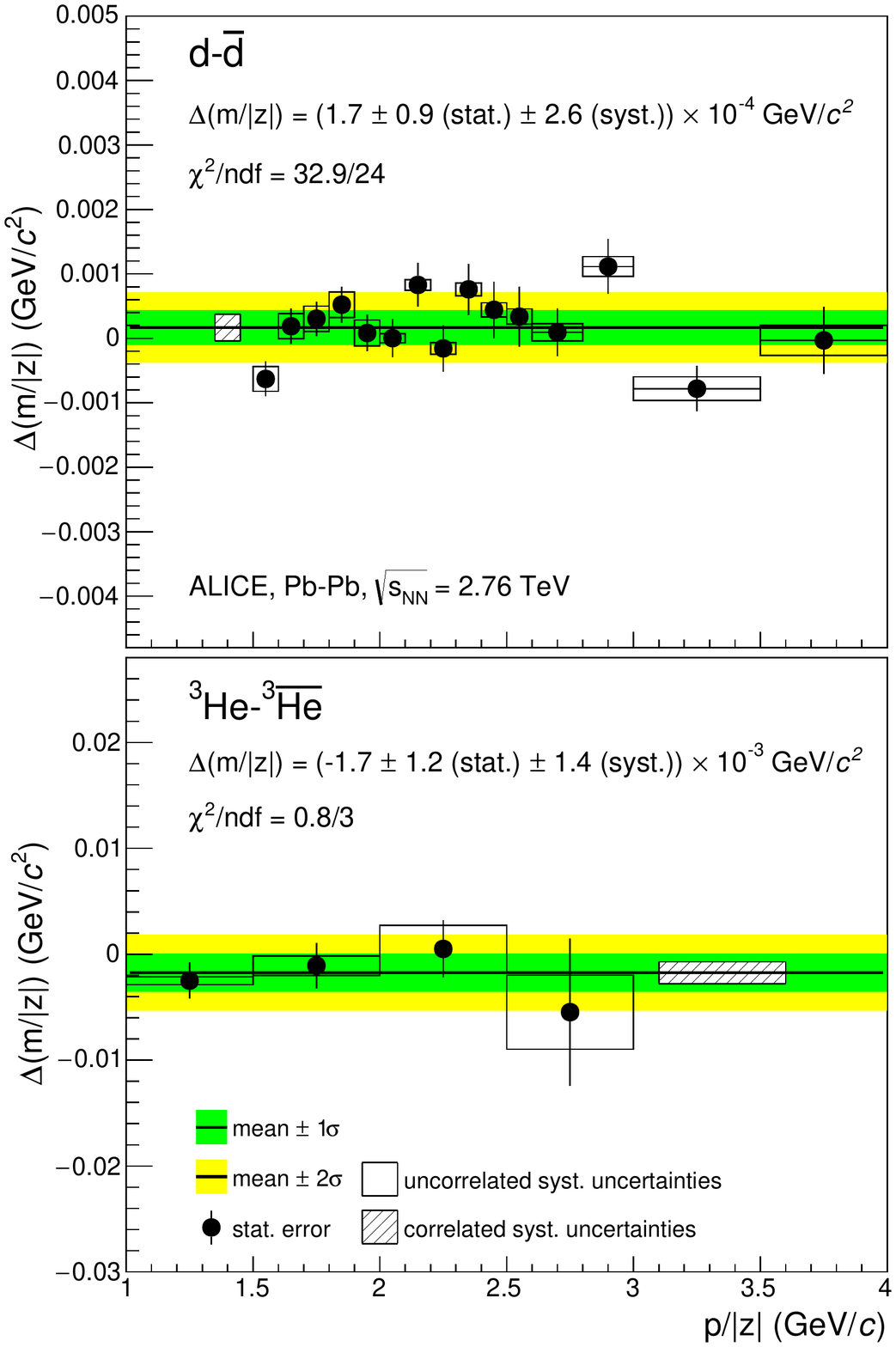}
\caption{The \deutns-\dbar (top) and \Hens-\Hebar (bottom) mass-over-charge ratio
difference measurements as a function of the particle rigidity.
Vertical bars and open boxes show the
statistical and the uncorrelated systematic uncertainties (standard deviations), respectively.
Both are taken into account to extract the
combined result in the full rigidity range,
together with the
correlated systematic uncertainty, which is shown as a box with tilted lines.
Also shown are the $1\sigma$ and $2\sigma$ bands around the central value,
where $\sigma$ is the sum in quadrature of the statistical and systematic uncertainties.
}
\label{fig:massDiff}
\end{center}
\end{figure}

The (anti-)deuteron and \mbox{(anti-)\He} masses are measured as the
peak position of the fitting curves of the mass-squared distribution.
The mass-over-charge ratio differences between the deuteron and \He and their respective anti-particle are then
evaluated as a function of the rigidity of the track, as shown in Fig.~\ref{fig:massDiff}.
The measurements in the individual rigidity intervals are
combined, taking into account statistical and systematic uncertainties (correlated
and uncorrelated), and the final result is shown in the same figure with one and two standard deviation uncertainty bands. 
The measured mass-over-charge ratio differences are
\begin{eqnarray} 
\Delta \mu_{\rm{d\bar{d}}} = \rm{[1.7 \pm 0.9 (stat.) \pm 2.6 (syst.)] \times 10^{-4}~GeV/{\it c}^2} \,, \label{eq:massValues_0}\\
\Delta \mu_{\rm{^{3}He ^{3}\overline{He}}} = \rm{[-1.7 \pm 1.2 (stat.) \pm 1.4 (syst.)] \times  10^{-3}~GeV/{\it c}^2} \,,
\label{eq:massValues}
\end{eqnarray}
corresponding to
\begin{eqnarray} 
\frac{\Delta \mu_{\rm{d}\rm{\overline{d}}}}{\mu_{\rm{d}}} = [0.9 \pm 0.5 (\rm{stat.}) \pm 1.4 (syst.)] \times 10^{-4} \,,\\
\frac{\Delta \mu_{\rm{^{3}He}\rm{^{3}\overline{He}}}}{\mu_{\rm{^{3}He}}} = [-1.2 \pm 0.9 (\rm{stat.}) \pm 1.0 (\rm{syst.})] \times 10^{-3} \,,
\label{eq:massPrecisions}
\end{eqnarray}
where $\mu_{\rm{d}}$ and $\mu_{\rm{^3He}}$ are the values recommended by CODATA\cite{CODATA}.
The mass-over-charge differences are compatible with zero within the estimated uncertainties,
in agreement with CPT invariance expectations. 

\begin{figure}
\begin{center}
\includegraphics[width=1.0\textwidth]{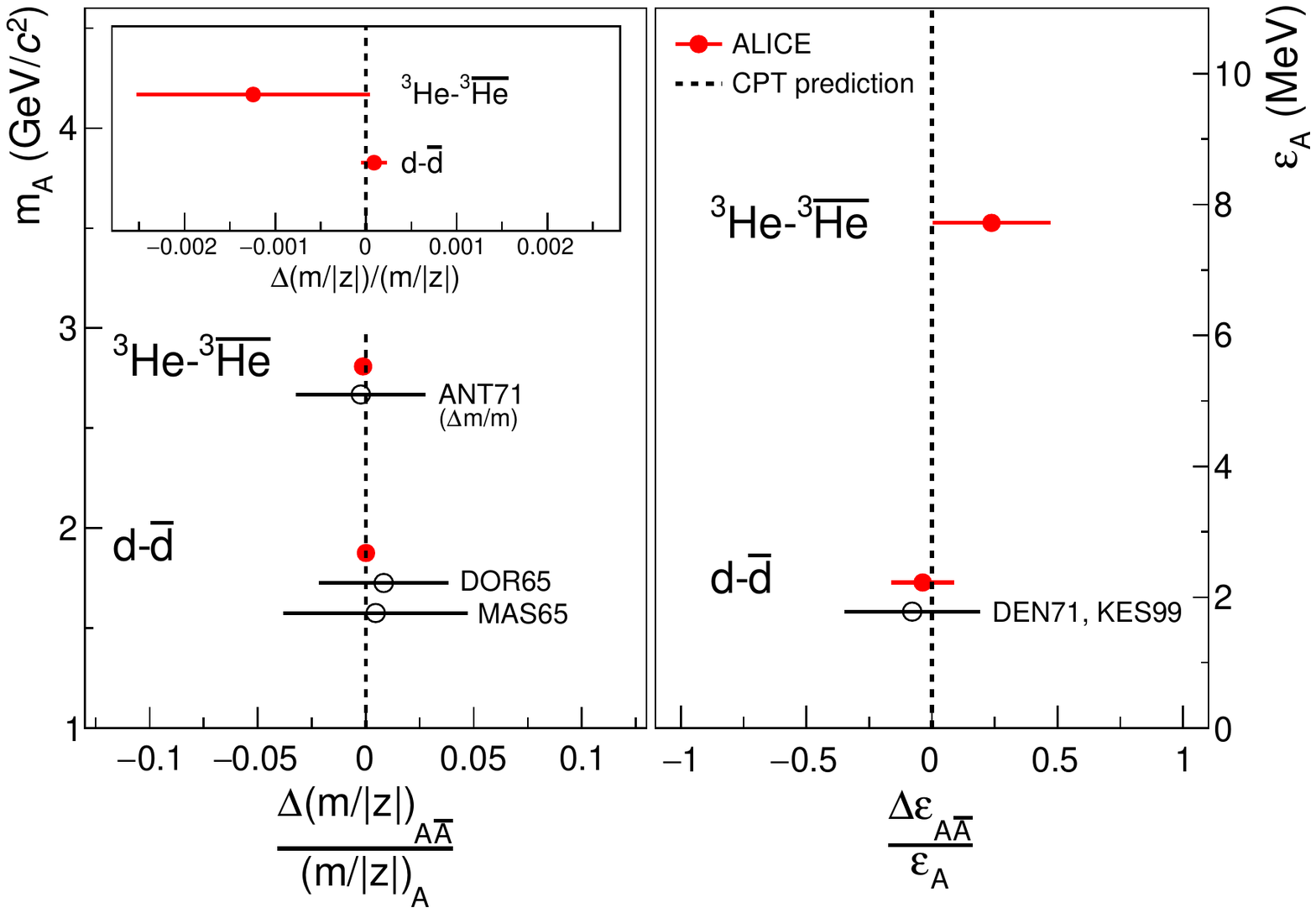}
\caption{The ALICE measurements for \deutns-\dbar and \Hens-\Hebar
mass-over-charge ratio differences compared with CPT invariance expectation (dotted lines) and
existing mass measurements MAS65\cite{dbarMassZick}, DOR65\cite{dbarMass} and ANT71\cite{he3Mass} (left panel).
The inset shows the ALICE results on a finer
\mbox{$\Delta(m/z)/(m/z)$} scale.
The right panel shows our determination of the binding energy differences
compared with direct measurements from DEN71\cite{bindingDbar} and KES99\cite{binddeut}.
Error bars represent the sum in quadrature of the statistical and systematic uncertainties
(standard deviations).
} 
\label{fig:massComparisons}
\end{center}
\end{figure}

Given that$z_{\bar{\rm{d}}} = - z_{\rm{d}}$ and $z_{^{3}\overline{\rm{He}}} = - z_{^{3}\rm{He}}$ as for the
proton and anti-proton\cite{pbarMass,TRAP},
the mass-over-charge differences in Eq.~\ref{eq:massValues_0} and Eq.~\ref{eq:massValues}
and the measurement of the mass differences between proton and anti-proton\cite{pbarMass,TRAP}
and between neutron and anti-neutron\cite{nbarMass,nbarMassErrata} can be used to derive the relative binding energy differences between the two studied particle species.
We obtain
\begin{eqnarray} 
\label{eq:deutBindLimit}
  \frac{\Delta \varepsilon_{\rm{d\bar{d}}}}{\varepsilon_{\rm{d}}} = -0.04 \pm 0.05~(\rm{stat.}) \pm 0.12~(\rm{syst.}) \,,\\
  \frac{\Delta \varepsilon_{\rm{^{3}He ^{3}\overline{He}}}}{\varepsilon_{\rm{^{3}He}}} = 0.24 \pm 0.16~(\rm{stat.}) \pm 
0.18~(\rm{syst.}) \,,
\label{eq:bindeHeLimit}
\end{eqnarray}
where $\varepsilon_{A} = Zm_{\rm{p}}+ (A-Z)m_{\rm{n}}-m_{A}$, being $m_{\rm{p}}$ and $m_{\rm{n}}$ the proton and the neutron mass values
recommended by PDG\cite{PDG} and $m_{A}$ the mass value of the nucleus with atomic number $Z$ and mass number $A$, recommended by CODATA\cite{CODATA}.
This quantity allows one to explicitly isolate possible violations of the CPT symmetry in the
(anti-)nucleon interaction from the one
connected to the (anti-)nucleon masses, the latter being constrained with a precision of $7 \times 10^{-10}$ for the proton/anti-proton system\cite{pbarMass,TRAP}.
Our results and the comparisons with previous mass difference measurements for (\deutns-\dbarns)\cite{dbarMassZick,dbarMass} and
(\Hens-\Hebarns)\cite{he3Mass}, as well as binding energy measurements for (\deutns-\dbarns)\cite{bindingDbar,binddeut}
are reported in Fig.~\ref{fig:massComparisons}.

We have shown that
the copious production of (anti-)nuclei in relativistic heavy-ion collisions
at the LHC represents a unique opportunity to test the CPT invariance of nucleon-nucleon interaction using light nuclei.
In particular, we have measured the mass-over-charge ratio differences for deuteron and \Hens.
The values are compatible, within uncertainties, with zero and represent a CPT invariance test in systems
bound by nuclear forces.
The results reported here (Fig.~\ref{fig:massComparisons}, left) represent the highest precision
direct measurements of mass differences in the sector of nuclei and they
improve by one to two orders of magnitude analogous results originally
obtained more than 40 years ago\cite{dbarMassZick,dbarMass,he3Mass}, and precisely 50 years ago for the anti-deuteron\cite{dbarMassZick,dbarMass}.
Remarkably such an improvement is reached in an experiment which is not
specifically dedicated to test the CPT invariance in nuclear systems.
In the forthcoming years the increase in luminosity and center-of-mass energy
at the LHC will allow to push forward the sensitivity of these measurements,
and possibly to extend the study to (anti-)${\rm ^4He}$.
Given the equivalence between mass and binding energy differences, our results also improve (Fig.~\ref{fig:massComparisons}, right) by a factor two the constraints on
CPT invariance inferred by existing measurements\cite{bindingDbar,binddeut} in the (anti-)deuteron system.
The binding energy difference has been determined for the first time
in the case of (anti-)\Hens, with a relative precision comparable to the one obtained in the (anti-)deuteron system.

\newpage
\newenvironment{acknowledgement}{\relax}{\relax}
\begin{acknowledgement}
\section*{Acknowledgements}
The ALICE Collaboration would like to thank all its engineers and technicians for their invaluable contributions to the construction of the experiment and the CERN accelerator teams for the outstanding performance of the LHC complex.
The ALICE Collaboration gratefully acknowledges the resources and support provided by all Grid centres and the Worldwide LHC Computing Grid (WLCG) collaboration.
The ALICE Collaboration acknowledges the following funding agencies for their support in building and
running the ALICE detector:
State Committee of Science,  World Federation of Scientists (WFS)
and Swiss Fonds Kidagan, Armenia,
Conselho Nacional de Desenvolvimento Cient\'{\i}fico e Tecnol\'{o}gico (CNPq), Financiadora de Estudos e Projetos (FINEP),
Funda\c{c}\~{a}o de Amparo \`{a} Pesquisa do Estado de S\~{a}o Paulo (FAPESP);
National Natural Science Foundation of China (NSFC), the Chinese Ministry of Education (CMOE)
and the Ministry of Science and Technology of China (MSTC);
Ministry of Education and Youth of the Czech Republic;
Danish Natural Science Research Council, the Carlsberg Foundation and the Danish National Research Foundation;
The European Research Council under the European Community's Seventh Framework Programme;
Helsinki Institute of Physics and the Academy of Finland;
French CNRS-IN2P3, the `Region Pays de Loire', `Region Alsace', `Region Auvergne' and CEA, France;
German Bundesministerium fur Bildung, Wissenschaft, Forschung und Technologie (BMBF) and the Helmholtz Association;
General Secretariat for Research and Technology, Ministry of
Development, Greece;
Hungarian Orszagos Tudomanyos Kutatasi Alappgrammok (OTKA) and National Office for Research and Technology (NKTH);
Department of Atomic Energy and Department of Science and Technology of the Government of India;
Istituto Nazionale di Fisica Nucleare (INFN) and Centro Fermi -
Museo Storico della Fisica e Centro Studi e Ricerche "Enrico
Fermi", Italy;
MEXT Grant-in-Aid for Specially Promoted Research, Ja\-pan;
Joint Institute for Nuclear Research, Dubna;
National Research Foundation of Korea (NRF);
Consejo Nacional de Cienca y Tecnologia (CONACYT), Direccion General de Asuntos del Personal Academico(DGAPA), M\'{e}xico, :Amerique Latine Formation academique – European Commission(ALFA-EC) and the EPLANET Program
(European Particle Physics Latin American Network)
Stichting voor Fundamenteel Onderzoek der Materie (FOM) and the Nederlandse Organisatie voor Wetenschappelijk Onderzoek (NWO), Netherlands;
Research Council of Norway (NFR);
National Science Centre, Poland;
Ministry of National Education/Institute for Atomic Physics and Consiliul Naţional al Cercetării Ştiinţifice - Executive Agency for Higher Education Research Development and Innovation Funding (CNCS-UEFISCDI) - Romania;
Ministry of Education and Science of Russian Federation, Russian
Academy of Sciences, Russian Federal Agency of Atomic Energy,
Russian Federal Agency for Science and Innovations and The Russian
Foundation for Basic Research;
Ministry of Education of Slovakia;
Department of Science and Technology, South Africa;
Centro de Investigaciones Energeticas, Medioambientales y Tecnologicas (CIEMAT), E-Infrastructure shared between Europe and Latin America (EELA), Ministerio de Econom\'{i}a y Competitividad (MINECO) of Spain, Xunta de Galicia (Conseller\'{\i}a de Educaci\'{o}n),
Centro de Aplicaciones Tecnológicas y Desarrollo Nuclear (CEA\-DEN), Cubaenerg\'{\i}a, Cuba, and IAEA (International Atomic Energy Agency);
Swedish Research Council (VR) and Knut $\&$ Alice Wallenberg
Foundation (KAW);
Ukraine Ministry of Education and Science;
United Kingdom Science and Technology Facilities Council (STFC);
The United States Department of Energy, the United States National
Science Foundation, the State of Texas, and the State of Ohio;
Ministry of Science, Education and Sports of Croatia and  Unity through Knowledge Fund, Croatia.
Council of Scientific and Industrial Research (CSIR), New Delhi, India
\end{acknowledgement}
\newpage


\begin{thebibliography}{30}
\bibitem{pbarMass} {Hori, M. \textit{et al.} Two-photon laser spectroscopy of antiprotonic helium and the antiproton-to-electron mass ratio. \textit{Nature} {\bf 475}, 484--488 (2011).}
\bibitem{TRAP} {Gabrielse, G. \textit{et al.} Precision Mass Spectroscopy of the Antiproton and Proton Using Simultaneously Trapped Particles. \textit{Phys. Rev. Lett.} {\bf 82}, 3198--3201 (1999).}
\bibitem{NuclForces} {van Kolck, U. Effective field theory of nuclear forces. \textit{Prog. Part. Nucl. Phys.} {\bf 43}, 337--418 (1999).}
\bibitem{ALICE} {Aamodt, K. \textit{et al.} ALICE Collaboration. The ALICE experiment at the CERN LHC. \textit{JINST} {\bf 3}, S08002 (2008).}
\bibitem{CPTtheorem_1}{L\"{u}ders, G. On the Equivalence of Invariance under Time Reversal and under Particle-Antiparticle Conjugation for Relativistic Field Theories. \textit{Kong. Dan. Vid. Sel. Mat. Fys. Med.} {\bf 28N5}, 1--17 (1954).}
\bibitem{CPTtheorem_2}{Pauli, W. Exclusion Principle, Lorentz Group and Reflection of Space-Time and Charge. \textit{Niels Bohr and the Development of Physics}, ed. by Pauli, W. (Pergamon Press, New York), 30--51 (1955).}
\bibitem{STARprod} {Agakishiev, H. \textit{et al.} STAR Collaboration. Observation of the antimatter helium-4 nucleus. \textit{Nature} {\bf 473}, 353--356 (2011).}%
\bibitem{STAR} {STAR Collaboration (Harris J. W. for the collaboration). The STAR experiment at the relativistic heavy ion collider. \textit{Nucl. Phys. A} {\bf 566}, 277C--285C (1994).}
\bibitem{PHENIX} {Gregory J. C. \textit{et al.} PHENIX Collaboration. PHENIX Experiment at RHIC. \textit{Nucl. Phys. A} {\bf 566}, 287C--298C (1994).}
\bibitem{posmass} {Fee, M. S. \textit{et al.} Measurement of the positronium $1~^3S_{1}$--$2~^3S_{1}$ interval by continuous-wave two-photon excitation. \textit{Phys. Rev. A} {\bf 48} 192--219 (1993).}
\bibitem{gFactor} {Van Dyck, R. S. Jr., Schwinberg P. B. and Dehmelt H. G. New High-Precision Comparison of Electron and Positron g Factors. \textit{Phys. Rev. Lett.} {\bf 59}, 26--29 (1987).} 
\bibitem{dm_bosons} {Abe, F. \textit{et al.} CDF Collaboration. A Measurement of the $W$-boson mass. \textit{Phys. Rev. Lett.} {\bf 65}, 2243--2246 (1990).}
\bibitem{ALPHA2012} {Amole, C. \textit{et al.} Resonant quantum transitions in trapped antihydrogen atoms. \textit{Nature} {\bf 483}, 439--443 (2012).}
\bibitem{Hbarcharge} {Amole, C. \textit{et al.} An experimental limit on the charge of antihydrogen. \textit{Nature Commun.} {\bf 5}, 3955 (2014).}
\bibitem{nbarMass} {Cresti, M., Pasquali, G., Peruzzo, L., Pinori, C. and Sartori, G. Measurement Of The Anti-neutron Mass. \textit{Phys. Lett. B} {\bf 177}, 206--210 (1986).}
\bibitem{nbarMassErrata} {Cresti, M., Pasquali, G., Peruzzo, L., Pinori, C. and Sartori, G. Errata. \textit{Phys. Lett. B} {\bf 200}, 587--588 (1988).}
\bibitem{ATRAP} {Di Sciacca, J. \textit{et al.} ATRAP Collaboration. One-Particle Measurement of the Antiproton Magnetic Moment. \textit{Phys. Rev. Lett.} {\bf 110}, 130801 (2013).}
\bibitem{k0CPT} {Ambrosino, F. \textit{et al.} KLOE Collaboration. Determination of CP and CPT violation parameters in the neutral kaon system using the Bell-Steinberger relation and data from the KLOE experiment. \textit{JHEP} {\bf 0612}, 011 (2006).}
\bibitem{sme} {Kosteleck\'y, V. A., Russel, N. Data tables for Lorentz and \textit{CPT} violation. \textit{Rev. Mod. Phys} {\bf 83}, 11--31 (2011).}
\bibitem{ALICEperf} {Abelev, B. I. \textit{et al.} ALICE Collaboration. Performance of the ALICE experiment at the CERN LHC. \textit{Int. J. Mod. Phys. A} {\bf 29}, 1430044 (2014).}
\bibitem{ITSperf} {Aamodt, K. \textit{et al.} ALICE Collaboration. Alignment of the ALICE Inner Tracking System with cosmic-ray tracks. \textit{JINST} {\bf 5}, P03003 (2010).}
\bibitem{TPCperf} {Alme, J. \textit{et al.} The ALICE TPC, a large 3-dimensional tracking device with fast readout for ultra-high multiplicity events. \textit{Nucl. Instrum. Meth. A} {\bf 622}, 316--367 (2010).}
\bibitem{TOFperf} {Akindinov, A. \textit{et al.} Performance of the ALICE Time-Of-Flight detector at the LHC. \textit{Eur. Phys. J. Plus} {\bf 128}, 44 (2013).}
\bibitem{PDG} {Olive, K. A. \textit{et al.} Particle Data Group Collaboration. Review of Particle Physics. \textit{Chin. Phys. C} {\bf 38}, 090001 (2014).}
\bibitem{CODATA} {Mohr, P. J., Taylor, B. N. and Newell, D. B. CODATA Recommended Values of the Fundamental Physical Constants: 2010. \textit{Rev. Mod. Phys.} {\bf 84}, 1527--1605 (2012).}
\bibitem{dbarMassZick} {Massam, T., Muller, Th., Righini, B., Schneegans, M. and Zichichi, A. Experimental observation of antideuteron production. \textit{Nuovo Cimento} {\bf 39}, 10--14 (1965).}
\bibitem{dbarMass} {Dorfan, D. E., Eades, J., Lederman, L. M., Lee, W. and Ting, C. C. Observation of antideuterons. \textit{Phys. Rev. Lett.} {\bf 14}, 1003--1006 (1965).} 
\bibitem{he3Mass} {Antipov, Yu. M. \textit{et al.} Observation of antihelium-3. \textit{Nucl. Phys. B} {\bf 31}, 235--252 (1971).}
\bibitem{bindingDbar}{Denisov, S. P. \textit{et al.} Measurements of anti-deuteron absorption and stripping cross sections at the momentum 13.3 GeV/$c$. \textit{Nucl. Phys. B} {\bf 31}, 253--260 (1971).}
\bibitem{binddeut} {Kessler, E. G. Jr. \textit{et al.} The Deuteron Binding Energy and the Neutron Mass. \textit{Phys. Lett. A} {\bf 255}, 221--229 (1999).}
\end{thebibliography}

\newpage
\appendix
\section{The ALICE Collaboration}
\label{app:collab}



\begingroup
\small
\begin{flushleft}
J.~Adam\Irefn{org39}\And
D.~Adamov\'{a}\Irefn{org82}\And
M.M.~Aggarwal\Irefn{org86}\And
G.~Aglieri Rinella\Irefn{org36}\And
M.~Agnello\Irefn{org110}\And
N.~Agrawal\Irefn{org47}\And
Z.~Ahammed\Irefn{org130}\And
I.~Ahmed\Irefn{org16}\And
S.U.~Ahn\Irefn{org67}\And
I.~Aimo\Irefn{org93}\textsuperscript{,}\Irefn{org110}
% ...
\renewcommand\labelenumi{\textsuperscript{\theenumi}~}

\section*{Affiliation notes}
\renewcommand\theenumi{\roman{enumi}}
\begin{Authlist}
\item \Adef{0}Deceased
\item \Adef{idp5882640}{Also at: University of Kansas, Lawrence, Kansas, United States}
\end{Authlist}

\section*{Collaboration Institutes}
\renewcommand\theenumi{\arabic{enumi}~}
\begin{Authlist}

\item \Idef{org1}A.I. Alikhanyan National Science Laboratory (Yerevan Physics Institute) Foundation, Yerevan, Armenia
\item \Idef{org2}Benem\'{e}rita Universidad Aut\'{o}noma de Puebla, Puebla, Mexico
\item \Idef{org3}Bogolyubov Institute for Theoretical Physics, Kiev, Ukraine
\item \Idef{org4}Bose Institute, Department of Physics and Centre for Astroparticle Physics and Space Science (CAPSS), Kolkata, India
\item \Idef{org5}Budker Institute for Nuclear Physics, Novosibirsk, Russia
\item \Idef{org6}California Polytechnic State University, San Luis Obispo, California, United States
\item \Idef{org7}Central China Normal University, Wuhan, China
\item \Idef{org8}Centre de Calcul de l'IN2P3, Villeurbanne, France
\item \Idef{org9}Centro de Aplicaciones Tecnol\'{o}gicas y Desarrollo Nuclear (CEADEN), Havana, Cuba
\item \Idef{org10}Centro de Investigaciones Energ\'{e}ticas Medioambientales y Tecnol\'{o}gicas (CIEMAT), Madrid, Spain
\item \Idef{org11}Centro de Investigaci\'{o}n y de Estudios Avanzados (CINVESTAV), Mexico City and M\'{e}rida, Mexico
\item \Idef{org12}Centro Fermi - Museo Storico della Fisica e Centro Studi e Ricerche ``Enrico Fermi'', Rome, Italy
\item \Idef{org13}Chicago State University, Chicago, Illinois, USA
\item \Idef{org14}China Institute of Atomic Energy, Beijing, China
\item \Idef{org15}Commissariat \`{a} l'Energie Atomique, IRFU, Saclay, France
\item \Idef{org16}COMSATS Institute of Information Technology (CIIT), Islamabad, Pakistan
\item \Idef{org17}Departamento de F\'{\i}sica de Part\'{\i}culas and IGFAE, Universidad de Santiago de Compostela, Santiago de Compostela, Spain
\item \Idef{org18}Department of Physics and Technology, University of Bergen, Bergen, Norway
\item \Idef{org19}Department of Physics, Aligarh Muslim University, Aligarh, India
\item \Idef{org20}Department of Physics, Ohio State University, Columbus, Ohio, United States
\item \Idef{org21}Department of Physics, Sejong University, Seoul, South Korea
\item \Idef{org22}Department of Physics, University of Oslo, Oslo, Norway
\item \Idef{org23}Dipartimento di Elettrotecnica ed Elettronica del Politecnico, Bari, Italy
\item \Idef{org24}Dipartimento di Fisica dell'Universit\`{a} 'La Sapienza' and Sezione INFN Rome, Italy
\item \Idef{org25}Dipartimento di Fisica dell'Universit\`{a} and Sezione INFN, Cagliari, Italy
\item \Idef{org26}Dipartimento di Fisica dell'Universit\`{a} and Sezione INFN, Trieste, Italy
\item \Idef{org27}Dipartimento di Fisica dell'Universit\`{a} and Sezione INFN, Turin, Italy
\item \Idef{org28}Dipartimento di Fisica e Astronomia dell'Universit\`{a} and Sezione INFN, Bologna, Italy
\item \Idef{org29}Dipartimento di Fisica e Astronomia dell'Universit\`{a} and Sezione INFN, Catania, Italy
\item \Idef{org30}Dipartimento di Fisica e Astronomia dell'Universit\`{a} and Sezione INFN, Padova, Italy
\item \Idef{org31}Dipartimento di Fisica `E.R.~Caianiello' dell'Universit\`{a} and Gruppo Collegato INFN, Salerno, Italy
\item \Idef{org32}Dipartimento di Scienze e Innovazione Tecnologica dell'Universit\`{a} del  Piemonte Orientale and Gruppo Collegato INFN, Alessandria, Italy
\item \Idef{org33}Dipartimento Interateneo di Fisica `M.~Merlin' and Sezione INFN, Bari, Italy
\item \Idef{org34}Division of Experimental High Energy Physics, University of Lund, Lund, Sweden
\item \Idef{org35}Eberhard Karls Universit\"{a}t T\"{u}bingen, T\"{u}bingen, Germany
\item \Idef{org36}European Organization for Nuclear Research (CERN), Geneva, Switzerland
\item \Idef{org37}Faculty of Engineering, Bergen University College, Bergen, Norway
\item \Idef{org38}Faculty of Mathematics, Physics and Informatics, Comenius University, Bratislava, Slovakia
\item \Idef{org39}Faculty of Nuclear Sciences and Physical Engineering, Czech Technical University in Prague, Prague, Czech Republic
\item \Idef{org40}Faculty of Science, P.J.~\v{S}af\'{a}rik University, Ko\v{s}ice, Slovakia
\item \Idef{org41}Faculty of Technology, Buskerud and Vestfold University College, Vestfold, Norway
\item \Idef{org42}Frankfurt Institute for Advanced Studies, Johann Wolfgang Goethe-Universit\"{a}t Frankfurt, Frankfurt, Germany
\item \Idef{org43}Gangneung-Wonju National University, Gangneung, South Korea
\item \Idef{org44}Gauhati University, Department of Physics, Guwahati, India
\item \Idef{org45}Helsinki Institute of Physics (HIP), Helsinki, Finland
\item \Idef{org46}Hiroshima University, Hiroshima, Japan
\item \Idef{org47}Indian Institute of Technology Bombay (IIT), Mumbai, India
\item \Idef{org48}Indian Institute of Technology Indore, Indore (IITI), India
\item \Idef{org49}Inha University, Incheon, South Korea
\item \Idef{org50}Institut de Physique Nucl\'eaire d'Orsay (IPNO), Universit\'e Paris-Sud, CNRS-IN2P3, Orsay, France
\item \Idef{org51}Institut f\"{u}r Informatik, Johann Wolfgang Goethe-Universit\"{a}t Frankfurt, Frankfurt, Germany
\item \Idef{org52}Institut f\"{u}r Kernphysik, Johann Wolfgang Goethe-Universit\"{a}t Frankfurt, Frankfurt, Germany
\item \Idef{org53}Institut f\"{u}r Kernphysik, Westf\"{a}lische Wilhelms-Universit\"{a}t M\"{u}nster, M\"{u}nster, Germany
\item \Idef{org54}Institut Pluridisciplinaire Hubert Curien (IPHC), Universit\'{e} de Strasbourg, CNRS-IN2P3, Strasbourg, France
\item \Idef{org55}Institute for Nuclear Research, Academy of Sciences, Moscow, Russia
\item \Idef{org56}Institute for Subatomic Physics of Utrecht University, Utrecht, Netherlands
\item \Idef{org57}Institute for Theoretical and Experimental Physics, Moscow, Russia
\item \Idef{org58}Institute of Experimental Physics, Slovak Academy of Sciences, Ko\v{s}ice, Slovakia
\item \Idef{org59}Institute of Physics, Academy of Sciences of the Czech Republic, Prague, Czech Republic
\item \Idef{org60}Institute of Physics, Bhubaneswar, India
\item \Idef{org61}Institute of Space Science (ISS), Bucharest, Romania
\item \Idef{org62}Instituto de Ciencias Nucleares, Universidad Nacional Aut\'{o}noma de M\'{e}xico, Mexico City, Mexico
\item \Idef{org63}Instituto de F\'{\i}sica, Universidad Nacional Aut\'{o}noma de M\'{e}xico, Mexico City, Mexico
\item \Idef{org64}iThemba LABS, National Research Foundation, Somerset West, South Africa
\item \Idef{org65}Joint Institute for Nuclear Research (JINR), Dubna, Russia
\item \Idef{org66}Konkuk University, Seoul, South Korea
\item \Idef{org67}Korea Institute of Science and Technology Information, Daejeon, South Korea
\item \Idef{org68}KTO Karatay University, Konya, Turkey
\item \Idef{org69}Laboratoire de Physique Corpusculaire (LPC), Clermont Universit\'{e}, Universit\'{e} Blaise Pascal, CNRS--IN2P3, Clermont-Ferrand, France
\item \Idef{org70}Laboratoire de Physique Subatomique et de Cosmologie, Universit\'{e} Grenoble-Alpes, CNRS-IN2P3, Grenoble, France
\item \Idef{org71}Laboratori Nazionali di Frascati, INFN, Frascati, Italy
\item \Idef{org72}Laboratori Nazionali di Legnaro, INFN, Legnaro, Italy
\item \Idef{org73}Lawrence Berkeley National Laboratory, Berkeley, California, United States
\item \Idef{org74}Lawrence Livermore National Laboratory, Livermore, California, United States
\item \Idef{org75}Moscow Engineering Physics Institute, Moscow, Russia
\item \Idef{org76}National Centre for Nuclear Studies, Warsaw, Poland
\item \Idef{org77}National Institute for Physics and Nuclear Engineering, Bucharest, Romania
\item \Idef{org78}National Institute of Science Education and Research, Bhubaneswar, India
\item \Idef{org79}Niels Bohr Institute, University of Copenhagen, Copenhagen, Denmark
\item \Idef{org80}Nikhef, National Institute for Subatomic Physics, Amsterdam, Netherlands
\item \Idef{org81}Nuclear Physics Group, STFC Daresbury Laboratory, Daresbury, United Kingdom
\item \Idef{org82}Nuclear Physics Institute, Academy of Sciences of the Czech Republic, \v{R}e\v{z} u Prahy, Czech Republic
\item \Idef{org83}Oak Ridge National Laboratory, Oak Ridge, Tennessee, United States
\item \Idef{org84}Petersburg Nuclear Physics Institute, Gatchina, Russia
\item \Idef{org85}Physics Department, Creighton University, Omaha, Nebraska, United States
\item \Idef{org86}Physics Department, Panjab University, Chandigarh, India
\item \Idef{org87}Physics Department, University of Athens, Athens, Greece
\item \Idef{org88}Physics Department, University of Cape Town, Cape Town, South Africa
\item \Idef{org89}Physics Department, University of Jammu, Jammu, India
\item \Idef{org90}Physics Department, University of Rajasthan, Jaipur, India
\item \Idef{org91}Physik Department, Technische Universit\"{a}t M\"{u}nchen, Munich, Germany
\item \Idef{org92}Physikalisches Institut, Ruprecht-Karls-Universit\"{a}t Heidelberg, Heidelberg, Germany
\item \Idef{org93}Politecnico di Torino, Turin, Italy
\item \Idef{org94}Purdue University, West Lafayette, Indiana, United States
\item \Idef{org95}Pusan National University, Pusan, South Korea
\item \Idef{org96}Research Division and ExtreMe Matter Institute EMMI, GSI Helmholtzzentrum f\"ur Schwerionenforschung, Darmstadt, Germany
\item \Idef{org97}Rudjer Bo\v{s}kovi\'{c} Institute, Zagreb, Croatia
\item \Idef{org98}Russian Federal Nuclear Center (VNIIEF), Sarov, Russia
\item \Idef{org99}Russian Research Centre Kurchatov Institute, Moscow, Russia
\item \Idef{org100}Saha Institute of Nuclear Physics, Kolkata, India
\item \Idef{org101}School of Physics and Astronomy, University of Birmingham, Birmingham, United Kingdom
\item \Idef{org102}Secci\'{o}n F\'{\i}sica, Departamento de Ciencias, Pontificia Universidad Cat\'{o}lica del Per\'{u}, Lima, Peru
\item \Idef{org103}Sezione INFN, Bari, Italy
\item \Idef{org104}Sezione INFN, Bologna, Italy
\item \Idef{org105}Sezione INFN, Cagliari, Italy
\item \Idef{org106}Sezione INFN, Catania, Italy
\item \Idef{org107}Sezione INFN, Padova, Italy
\item \Idef{org108}Sezione INFN, Rome, Italy
\item \Idef{org109}Sezione INFN, Trieste, Italy
\item \Idef{org110}Sezione INFN, Turin, Italy
\item \Idef{org111}SSC IHEP of NRC Kurchatov institute, Protvino, Russia
\item \Idef{org112}SUBATECH, Ecole des Mines de Nantes, Universit\'{e} de Nantes, CNRS-IN2P3, Nantes, France
\item \Idef{org113}Suranaree University of Technology, Nakhon Ratchasima, Thailand
\item \Idef{org114}Technical University of Split FESB, Split, Croatia
\item \Idef{org115}The Henryk Niewodniczanski Institute of Nuclear Physics, Polish Academy of Sciences, Cracow, Poland
\item \Idef{org116}The University of Texas at Austin, Physics Department, Austin, Texas, USA
\item \Idef{org117}Universidad Aut\'{o}noma de Sinaloa, Culiac\'{a}n, Mexico
\item \Idef{org118}Universidade de S\~{a}o Paulo (USP), S\~{a}o Paulo, Brazil
\item \Idef{org119}Universidade Estadual de Campinas (UNICAMP), Campinas, Brazil
\item \Idef{org120}University of Houston, Houston, Texas, United States
\item \Idef{org121}University of Jyv\"{a}skyl\"{a}, Jyv\"{a}skyl\"{a}, Finland
\item \Idef{org122}University of Liverpool, Liverpool, United Kingdom
\item \Idef{org123}University of Tennessee, Knoxville, Tennessee, United States
\item \Idef{org124}University of the Witwatersrand, Johannesburg, South Africa
\item \Idef{org125}University of Tokyo, Tokyo, Japan
\item \Idef{org126}University of Tsukuba, Tsukuba, Japan
\item \Idef{org127}University of Zagreb, Zagreb, Croatia
\item \Idef{org128}Universit\'{e} de Lyon, Universit\'{e} Lyon 1, CNRS/IN2P3, IPN-Lyon, Villeurbanne, France
\item \Idef{org129}V.~Fock Institute for Physics, St. Petersburg State University, St. Petersburg, Russia
\item \Idef{org130}Variable Energy Cyclotron Centre, Kolkata, India
\item \Idef{org131}Vin\v{c}a Institute of Nuclear Sciences, Belgrade, Serbia
\item \Idef{org132}Warsaw University of Technology, Warsaw, Poland
\item \Idef{org133}Wayne State University, Detroit, Michigan, United States
\item \Idef{org134}Wigner Research Centre for Physics, Hungarian Academy of Sciences, Budapest, Hungary
\item \Idef{org135}Yale University, New Haven, Connecticut, United States
\item \Idef{org136}Yonsei University, Seoul, South Korea
\item \Idef{org137}Zentrum f\"{u}r Technologietransfer und Telekommunikation (ZTT), Fachhochschule Worms, Worms, Germany
\end{Authlist}
\endgroup

\end{document}